\documentclass[manuscript]{acmart}

\usepackage[utf8]{inputenc}
\usepackage{subcaption}

\usepackage[resetlabels]{multibib}
\newcites{game}{Ludography}
\nocitegame{*}
\newcommand{\citegameprefix}{G}
\usepackage{xparse}
\let\origcitegame\citegame
\RenewDocumentCommand{\citegame}{O{} O{} m}{%
  \renewcommand{\citenumfont}[1]{\citegameprefix##1}%
  \origcitegame[#1][#2]{#3}%
  \renewcommand{\citenumfont}[1]{##1}%
}
\newcommand{\citeg}[1]{\citegame{#1}} 

\AtBeginDocument{%
  }

\begin{document}

\title{Playing to Pay: Interplay of Monetization and Retention Strategies in Korean Mobile Gaming}


\author{HwiJoon Lee}
\affiliation{%
  \institution{Northeastern University}
  \city{Boston}
  \country{USA}}
\email{lee.hw@northeastern.edu}

\author{Kashif Imteyaz}
\affiliation{%
  \institution{Northeastern University}
  \city{Boston}
  \country{USA}}
\email{Imteyaz.k@northeastern.edu}

\author{Saiph Savage}
\affiliation{%
  \institution{Northeastern University}
  \city{Boston}
  \country{USA}}
\email{s.savage@northeastern.edu}


\begin{abstract}
Mobile gaming's global growth has introduced evolving monetization strategies, such as in-app purchases and ads, designed to boost revenue while maintaining player engagement. However, there is limited understanding of the scope and frequency of these strategies, particularly in mature markets like South Korea. To address this research gap, this study examines the monetization strategies used in the top 40 most popular Korean mobile games through direct gameplay observations and targeted video analyses. We identified the prevalence of specific strategies, including time-gated progression, Conflict-Driven Design, and social Dynamics, which are systematically categorized in our proposed framework for monetization. Our findings also highlight ethical concerns, including issues with transparency, probability disclosures, and the exploitation of competitive pressures—areas that remain poorly regulated. To address these challenges, we emphasize the need for stricter consumer protections, cross-regional research, and greater focus on protecting vulnerable populations to promote a more equitable and responsible gaming environment.
\end{abstract}

\ccsdesc[500]{Human-centered computing~User centered design}

\keywords{ Monetization strategies, Player retention, Mobile games, Gacha mechanics, Game design ethics, Korean mobile game market, Player spending behavior }


\maketitle
\section{INTRODUCTION}
Monetization strategies in mobile games have become a central aspect of game design, shaping both player behavior and revenue generation \cite{ElNasr13}. These strategies are methods used to maximize revenue while maintaining player engagement \cite{Nieborg16}. Examples include: \emph{In-App Purchases:} Players buy virtual items, currencies, upgrades, or customization options; \emph{Advertisements:} Revenue is generated through in-game ads, such as rewarded ads or banners; \emph{Loot Boxes or Gacha Systems:} Players pay for randomized items or rewards \cite{Nieborg16}. Over the past decade, the global mobile game market has expanded significantly, driven by rising smartphone usage and increasingly sophisticated monetization strategies \cite{Capcom24}. While early mobile games typically followed a ``premium'', pay-upfront model, developers have largely shifted to ``freemium'' approaches \cite{Karlsen22, Nieborg16}. These newer models incorporate in-game purchases, loot boxes, and advertisements \cite{Nieborg16, Hamari2020"Why}. Such techniques create emotional and financial investments, which can blur the line between entertainment and exploitation \cite{Tom20}.

Recent research has examined the psychological mechanisms underlying monetization strategies—such as ``loss aversion,'' ``the sunk cost fallacy,'' and ``scarcity effects''—to explain why players may feel compelled to spend more than they initially intended \cite{Ronayne2019Evaluating, Tom20}. For example, models like random reward systems have been linked to problematic player behaviors, including compulsive spending and addictive tendencies \cite{ElNasr13, King19, Dreier17}. Consequently, ethical concerns have grown regarding the potential exploitation of vulnerable groups and the promotion of harmful financial habits through these monetization strategies \cite{Freeman2022Pay}. Moreover, in regions like South Korea—where rapid technological adoption and a robust gaming culture have accelerated the prominence of gaming—these issues become especially pronounced \cite{Lee2020The, Jin2015Transformative}. Despite advanced infrastructures and regulatory frameworks, players in such markets still appear susceptible to exploitative tactics, leading to increasing consumer pushback \cite{Park2023Learnings}.

However, despite extensive discussion of monetization models in the existing literature \cite{ElNasr13, Freeman2022Pay, Zendle2020Paying}, there remains a limited understanding of the range of strategies employed in mobile games and how frequently they are implemented. This gap is particularly significant in mature gaming markets, such as South Korea, where these practices may pose unique behavioral and ethical challenges.  

To address this research gap, our paper systematically examines the monetization strategies used in Korean mobile games and investigates their potential influence on player behavior. Through thematic analysis, we identify and define a wide range of monetization tactics and introduce a new terminology framework for categorizing them.  Our findings reveal how mobile game developers combine various monetization strategies to create immediate spending incentives while maintaining sustained player engagement, ultimately maximizing revenue. By uncovering these evolving design practices and their associated risks, this study provides a more comprehensive understanding of mobile game monetization and its ethical implications in a global context.

\section{METHODOLOGY}
We selected a sample of 40 mobile games from the most popular titles in the Korean Google Play Store (as of October 2024), using the Google Play top-grossing rankings as a reference \cite{google_play_ranking}. The complete list of games included in our analysis is available in the Ludography section. Our study focused on examining the monetization strategies employed in these popular mobile games. To achieve this, we adopted methods similar to those used in prior research \cite{Poretski22, Toups19, wuertz2018awareness}. 

First, we engaged in direct gameplay for each title, playing 2–4 hours per game to collect data about the monetization strategies present in these games. Second, to capture monetization features that may only appear after extended play—particularly in Massively Multiplayer Online Role-Playing Games (MMORPGs)—we supplemented our analysis of the 13 MMORPG titles with at least five playthrough videos per game. These videos were specifically selected if they included moments of in-game purchasing, featured minimal external commentary or audience interaction, and provided a clear look at what led up to each purchase. By focusing on footage that highlighted both the purchasing act and its immediate triggers, we were able to observe monetization features in context, free from distracting commentary. This approach allowed us to capture monetization tactics that might not be evident in shorter gameplay sessions, offering a more comprehensive view of the strategies employed by game developers.

Following data collection, we conducted an iterative thematic analysis guided by inductive qualitative methods \cite{Braun06}. Our goal was to categorize the various monetization strategies present in these mobile games. During the initial familiarization phase, we identified key monetization moments from both direct gameplay and video observations, generating open codes to capture critical aspects of each strategy. These codes were then organized into broader categories by identifying patterns and relationships, leading to the identification of overarching design strategies that can influence player spending and retention. This thematic approach is consistent with methods used in related studies \cite{Poretski22, Toups19, wuertz2018awareness}, allowing us to systematically highlight recurring monetization techniques. Examples include tactics that create urgency or reduce perceived player effort, which were placed within a broader framework of practices that significantly shape player behavior and spending decisions.

\section{FINDINGS}
\begin{table}[h]
\centering
\renewcommand{\arraystretch}{1.2}
\begin{tabular}{p{3.5cm}p{8cm}r}
\toprule
\textbf{Strategy} & \textbf{Definition} & \textbf{Games (\#)} \\
\midrule
Daily Login & Provides rewards for consistent daily logins, encouraging habitual engagement. & 38 \\
Loot Box & Implements randomized reward systems that encourage spending for a chance to acquire rare items. & 37 \\
Battle Pass & Offers tiered rewards tied to progression, often encouraging spending to unlock additional or premium content. & 31 \\
Progress Boosters & Allows players to gain strength or progress faster, emphasizing early investment to unlock better rewards sooner. & 27 \\
Social Dynamics & Leverages multiplayer interactions to drive engagement and spending for cooperative goals. & 26 \\
Conflict-Driven Design & Encourages player spending through competitive dynamics to gain advantages over others for future competition. & 25 \\
First Purchase Benefits & Incentivizes the first in-game purchase with exclusive or highly rewarding bonuses. & 17 \\
Collection Gacha & Rewards incremental progress toward completing a set of items, with bonuses for full completion. & 12 \\
Time-Gated Progression & Restricts player advancement with time delays, which can be bypassed by spending money. & 9 \\
Ease of Play Features & Sells features that simplify gameplay, reduce manual effort, or enhance the overall player experience. & 9 \\
Complete Gacha & Requires players to collect a complete set of randomized items to unlock a final reward, often involving high investment. & 3 \\
\bottomrule
\end{tabular}
\caption{Instances of Monetization and Retention Strategies across Analyzed Games.}
\label{tab:strategies}
\end{table}

Through our analysis of mobile game design in the Korean market, we identified two primary categories of strategies that developers use to maximize revenue: \textbf{Monetization Strategies} and \textbf{Retention Strategies}. Monetization strategies focus on driving player spending by offering opportunities to accelerate progress, obtain rare rewards, or gain competitive advantages. On the other hand, retention strategies aim to keep players engaged over time, ensuring they continue to interact with the game and ultimately spend more. By combining high player spending with long-term engagement, these strategies work to optimize the financial success of mobile games. The following sections break down these findings into specific mechanics and psychological principles observed in the games analyzed.

\subsection{Monetization Strategies}
Here we present  the different monetizations strategies we identified in our study. 

\subsubsection{Ease of Play Monetization Strategy}
In mobile games, players may experience physical discomfort during prolonged play sessions \cite{Berolo11, Hanphitakphong2021Effect}. To address this, developers often implement “ease of play” mechanisms designed to minimize repetitive or time-consuming tasks within the game, as we observed during our analysis. However, our findings reveal that these mechanisms are frequently tied to monetization strategies, encouraging players to spend money to bypass inconveniences and enhance their ease of play. This aligns with prior research, which has shown that developers often exploit players’ perceived need for convenience to drive spending \cite{Petrovskaya22, Hamari2020}. Notably, we identified that 25\% of the games in our study incorporated this monetization mechanism.

An example of this type of monetization strategy is seen in Honkai Star Rail \citeg{honkaiStarRail}, where players can obtain stronger characters to automate gameplay more effectively. Some powerful units allow players to clear challenging stages using auto-play, reducing the need for manual engagement. However, once obtained, these characters require further investment in resources to enhance their abilities, driving players to engage in additional auto-play cycles to gather materials. This creates a feedback loop where convenience drives both spending and continued engagement, appealing to players seeking faster progression with minimal active participation.

\subsubsection{Monetization Strategies Leveraging Time-Gated Progression and Progress Boosters}
We observed that many games (25\% in our dataset) incorporated time-gated progression mechanisms, which limited player advancement by imposing delays. These games were often designed to provide players with two options: either wait for the progression timer to expire or bypass the delay through paid options. This monetization strategy thus capitalizes on players' desire to accelerate their progress, offering strategic purchase opportunities to advance more quickly.

\begin{figure}[h] 
    \centering
    \includegraphics[width=0.3\linewidth]{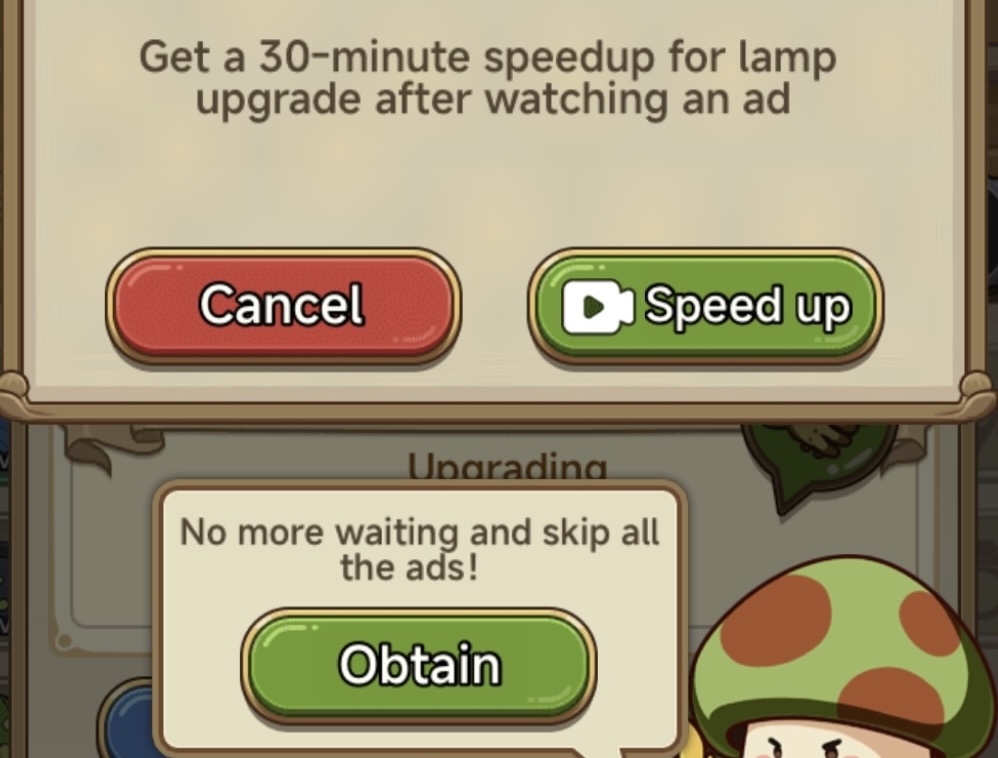}
    \caption{Ad-skipping option in Legend of Mushroom}
    \label{fig:legend-of-mushroom-ads}
\end{figure}

For example, \textit{Legend of Mushroom} \citeg{LegendOfMushroom} incorporates this monetization strategy (see Fig. \ref{fig:legend-of-mushroom-ads}). In this game, players can accelerate their daily progress by watching ads to gain temporary boosts. To avoid the inconvenience of repeatedly watching ads, the game provides an option to permanently skip ads for a fee. This feature is framed as a smart investment, offering consistent rewards that appeal to players who want to save time and minimize interruptions. By trading efficiency for money, this strategy effectively encourages players to make purchases to enhance their in-game performance \cite{Goldstein2014The}.

\subsubsection{Monetization Strategies with Randomized Reward Systems (Gacha mechanics)} 
Gacha mechanics, also referred to as loot boxes, are a type of monetization model commonly used in mobile games \cite{Zendle2019The}, where players acquire in-game items through randomized draws, typically by spending real money or in-game currency \cite{Toto12, King19}. The core appeal of gacha systems lies in the thrill of chance, as players aim to obtain rare and valuable rewards \cite{Zendle2019The}. However, this system can also be financially burdensome, as players are incentivized to spend repeatedly, driven by the possibility of acquiring coveted items \cite{King19}.  In our dataset, we observed that 92.5\% of the games incorporated at least one gacha mechanics as a monetization strategy. Additionally, we identified two advanced forms in which this monetization strategy was implemented:

\begin{figure}[h]
\centering
\includegraphics[width=0.40\linewidth]{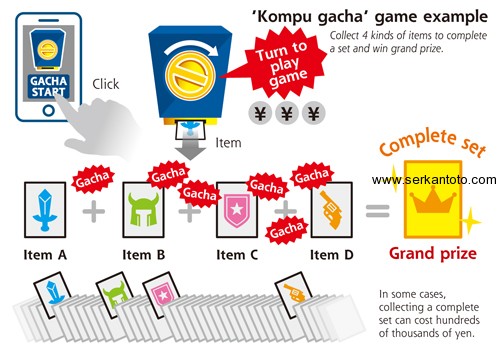}
\caption{An example of the Complete Gacha system \cite{Toto12}}
\label{fig:comp-gacha-explained}
\end{figure}

\textbf{Complete Gacha}  
In some gacha systems, players must collect a complete set of items to unlock a powerful reward, but the process often involves considerable risk and investment \cite{Toto12, Shibuya15, DeVere12}. These systems require players to obtain multiple specific parts, with each part typically having low odds of being obtained. Players must collect individual items through randomized gacha pulls. Only after all items are obtained can the player unlock the grand prize. This structure pushes players to continue spending, as incomplete sets hold little value, often making the system both financially demanding and frustrating \cite{Toto12}

Although many companies and developers in Japan have agreed to restrict the use of the Complete Gacha system due to its controversial nature \cite{Ernkvist16}, our analysis identified a few instances \citeg{lineage2m, lineagem, lineagew} of this monetization strategy in the Korean mobile market. In \textit{Lineage 2M}, players are required to collect 10 distinct parts to craft a high-tier item. Each part is obtained through randomized gacha pulls, meaning players often need to engage in repeated attempts to acquire the necessary items. The system requires players to collect all 10 parts to craft the final item; until every part is obtained, the progress made holds no value. This design encourages players to continue engaging with the system, driven by the desire to complete their investment and avoid leaving it unfinished.

\textbf{Collection Systems} present a subtle variation of the complete gacha mechanic. Each character or item within a collection holds its own value, providing incremental benefits even if the set remains incomplete. Unlike complete gacha, where obtaining all pieces is necessary to unlock all benefits, collection systems offer progressive bonuses as players gather more items. However, completing the set is still required to maximize the reward. This system benefits mobile game developers by allowing them to introduce new items periodically, motivating players to continue spending on gacha pulls to complete their collections. From our observation, this mechanic creates a continuous loop of engagement, encouraging players to invest further in order to achieve the ultimate set reward. Notably, the potential harm caused by this form of loot boxes to players has not been sufficiently investigated in existing researches.

\subsubsection{Monetization Strategies with Conflict-Driven Design}
In our analysis, we observed that conflict between players plays a significant role in encouraging in-game spending. Many of these games create environments where competition over limited resources or rewards intensifies the desire for players to gain an advantage. This competitive structure often incentivizes players to spend money to become more powerful or to stay ahead of others. 62.5\% of the games in our dataset incoprorated this strategy. 

For example, we found that in every Massively Multiplayer Online Role-Playing Games (MMORPGs), a genre where players control characters and interact with one another in a persistent virtual world \cite{Cole2007Social}, high-level bosses or rare resources are limited and serve as key milestones in the game. Players compete for the chance to defeat these bosses first for the rare and valuable rewards. This scarcity creates pressure for players to spend money on in-game items or boosts that allow them to grow stronger more quickly. This is particularly evident in the early stages of a game's lifecycle, where players feel the need to establish dominance by securing these rewards before others can.

We also observed that in-game communication systems, such as ``global or guild chat'', often intensified competition among players. These systems enable players to openly challenge one another, potentially escalating rivalries and increasing emotional investment in gaining a competitive edge. Additionally, the leaderboards of these mobile games also offer exclusive rewards for top players or guilds, which could further incentivize competition to maintain high rankings.  Research on competition in video games suggests that competitive modes enhance player satisfaction and immersion by fostering a strong flow experience \cite{Sepehr2017Understanding}. This heightened engagement may help explain why players in MMORPGs are more likely to invest in in-game purchases to sustain a competitive advantage.

\subsection{Retention Strategies}
We observed that monetization strategies in mobile games were often closely linked to retention strategies. As a result, we also examined common retention strategies present in our dataset.  In gaming, retention strategies frequently leverage psychological principles centered on a player's desire to avoid losses—whether these losses are anticipated in the future or related to prior investments \cite{ElNasr13}. These principles work together to encourage sustained engagement by making players feel that stopping play or spending would result in missed benefits or wasted past efforts \cite{ElNasr13}.

\subsubsection{Retention Strategies Leveraging Future Loss Aversion Mechanisms}
Loss aversion is a cognitive bias in which potential losses evoke stronger negative emotions than equivalent gains elicit positive emotions \cite{Tom20}. In gaming, this bias is reflected in retention strategies designed to incentivize daily engagement, such as daily login rewards or daily charging bonuses \cite{Tom20}. Missing out on these rewards can create feelings of loss, motivating players to engage with the game regularly \cite{Tom20}.

\textbf{Daily Login Rewards}  
Daily login rewards are the most popular strategy (used by 95\% of games) designed to establish habitual behavior. Players receive rewards for regular logins, which over time builds a routine of daily interaction with the game. Missing a day can feel like losing out on accumulated progress, encouraging players to log in daily. This habitual engagement sustains player activity and creates a foundation for monetization opportunities. By keeping players consistently involved, retention strategies like daily login rewards increase exposure to the monetization mechanisms. This synergy between retention and monetization ensures a continuous feedback loop, where sustained engagement fosters spending, and monetization mechanisms, in turn, incentivize players to remain active, maximizing revenue for the company.

\subsubsection{Retention Strategies Incorporating Protecting Past Investments (Sunk Cost Fallacy).}
The sunk cost fallacy is a psychological principle in which individuals continue investing time and resources into an activity to avoid feeling that their previous efforts were wasted \cite{Arkes1985The}. We found that mobile games employ this principle as a retention strategy by leveraging players' past investments—such as time spent progressing, acquiring items, or completing achievements—to encourage ongoing engagement.  This strategy can help to reinforce both emotional and financial commitment, as players are motivated to avoid losing the perceived value of their earlier efforts. Furthermore, we observed that this strategy manifested in two main ways:

\textbf{Encouraging First Purchases}  
Game designers often capitalize on the sunk cost fallacy by encouraging initial spending, which influences players to continue spending over time \cite{Liang2014The} Tactics such as one-time bonuses for first-time buyers lower the barrier to initial spending and create a perceived sense of value. Once players invest, they become more likely to stay engaged, motivated by the desire to build upon their initial investment.

\begin{figure}[h]
\centering
\includegraphics[width=0.7\linewidth]{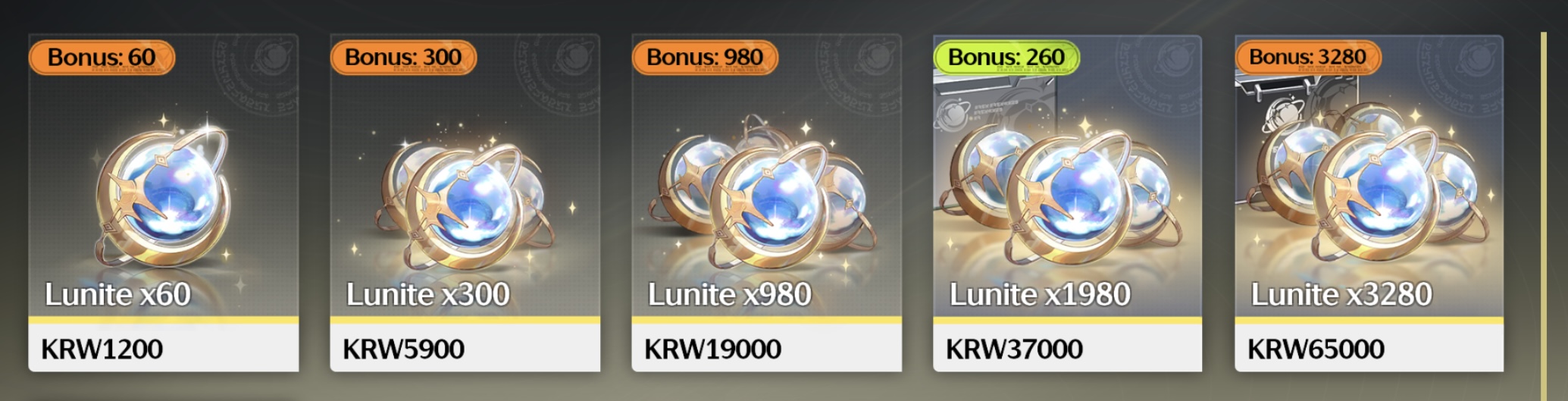}
\caption{First purchase bonuses in Wuthering Waves \citeg{wutheringWaves}, where players receive double value (orange) for initial purchases. After the first purchase, the bonus decreases to 13\% extra (green), encouraging initial spending with greater rewards.}
\label{fig:Wuthering_Waves}
\end{figure}

For example, as illustrated in Figure \ref{fig:Wuthering_Waves}, we observed that some games in our dataset offer double bonuses for first-time buyers, prominently marked in orange with a ``2x value'' label. Subsequent bonuses decrease significantly, offering only a 13\% bonus (marked in green). This approach incentivizes an initial purchase by providing a one-time, highly rewarding bonus. After making this initial investment, players are often encouraged to continue engaging with the game to avoid feeling that their prior expenditure was wasted.

\textbf{Social Dynamics}  
In competitive and cooperative settings, social dynamics also contribute to retention. Players often feel compelled to keep up with their guild members or peers, which can reinforce the sunk cost effect. In competitive MMORPGs, where guilds compete for limited rewards, players are driven to invest more time and resources to maintain standing within their social groups \cite{Hamari2017Why}. Unity’s recent report underscores the importance of multiplayer dynamics in modern game design, supporting the role of social influences on player engagement and spending \cite{Unity24}.

\section{DISCUSSION}
\subsection{Ethical Concerns}
The mobile game industry has achieved significant financial success, generating a global revenue of 136.2 billion U.S. dollars \cite{Capcom24}. However, despite its economic impact, the industry's aggressive monetization tactics—such as complete or collection gacha that were identified in our study—raise substantial ethical concerns due to insufficient regulatory oversight \cite{Zendle2020Paying, Park2023Learnings}.  As our results show, players are often exposed to psychologically exploitative designs. These designs frequently leverage psychological principles to incentivize spending \cite{Liang2014The, Nieborg16, ElNasr13}, yet there are limited consumer protections in place to mitigate these practices \cite{Xiao2020Regulating, Xiao2023What}. This lack of legal safeguards allows companies to continue employing monetization techniques that exploit cognitive biases and social pressures to drive revenue \cite{Xiao2020Regulating}.

A key ethical concern here is the lack of transparency in probability-based monetization mechanics like gacha systems \cite{Park2023Learnings, Xiao2023What}. In some cases, gaming companies have been found to provide misleading information about the probabilities of obtaining rare items , with the actual odds being lower than those publicly stated \cite{Park2023Learnings}. Since players lack access to the data needed to independently verify these probabilities, they have no reliable way to determine whether they are being offered fair chances or subjected to deceptive practices \cite {Park2023Learnings}. This information asymmetry places players at a significant disadvantage, preventing them from making informed purchasing decisions and leaving them vulnerable to potential exploitation. To address this issue, there is a growing call for legal interventions that mandate gaming companies to implement transparent and verifiable probability disclosures \cite{kimchang2024} For instance, regulations could require companies to maintain publicly accessible logs or third-party audits of probability data, enabling players to confirm the accuracy of advertised probabilities \cite{kimchang2024}. Without such measures, players will remain unable to substantiate their claims or challenge unfair practices, effectively leaving companies unchecked in their use of these tactics.

Another ethical concern is the use of conflict-driven game design, particularly in multiplayer formats like MMORPGs \cite{Cole2007Social, Freeman2022Pay}. As highlighted in our study, such designs can amplify competitive pressures, which in turn can drive monetization. In these environments, players often feel compelled to spend money to maintain their status or outperform rivals \cite{Freeman2022Pay}. As our research showed, mobile game developers often exploit this dynamic by offering in-game purchases that provide competitive advantages.  These ``pay-to-win'' mechanics create a high-stakes environment where players may feel pressured to make repeated purchases to remain competitive \cite{Freeman2022Pay}. Research indicates that such designs can blur the line between voluntary spending and compulsion \cite{Garea2023The}, especially when players are frequently reminded of the benefits obtained through spending \cite{Hamari2017Why, Zendle2020Paying}.

To mitigate these ethical issues, the industry must prioritize consumer protection and transparency in its monetization practices. Implementing enforceable regulations on probability disclosures and restricting predatory monetization tactics can help address the power imbalance between gaming companies and players \cite{kimchang2024}. Clear standards can prevent exploitative designs \cite{Xiao2020Regulating}, regulators can foster a fairer environment where players can enjoy games without facing undue psychological pressure or financial exploitation \cite{Xiao2020Regulating, Xiao2023What}. Ultimately, adopting these measures is essential to aligning the industry’s practices with ethical standards that respect players’ rights and well-being, ensuring that financial success does not come at the expense of player trust and fairness.

\subsection{Limitations}
This study employed a qualitative thematic analysis to examine monetization strategies in mobile games, identifying recurring design patterns. However, as an interpretive method, it introduces a level of subjectivity that may limit the generalizability of the findings.  The analysis also relied on gameplay and video observations but did not include direct feedback from players. Incorporating firsthand accounts could provide deeper insights into how monetization strategies influence engagement and spending. Future research could address this gap by conducting player interviews to better understand their perspectives on these strategies.  Additionally, while the study identified notable trends, it lacks empirical data to establish causal relationships. Future work should include experiments to expand upon these findings.

\subsection{Future Works and Conclusion}
Our study provides valuable insights into the monetization strategies of the Korean mobile game market, but it also highlights key opportunities for future research. Expanding the scope to include other regions could uncover how cultural and regulatory differences influence monetization strategies, enabling comparative studies to identify global patterns.  

Future research should also prioritize real player experiences through surveys, interviews, or experimental studies. Such approaches could provide deeper insights into the psychological and emotional effects of monetization on diverse player demographics, including vulnerable groups. Additionally, quantitative methods, such as large-scale surveys or behavioral tracking, could complement the trends identified in this study. Longitudinal studies could further examine long-term impacts, such as addiction and compulsive spending.  

In conclusion, our study highlights the interaction between game design and monetization strategies and their potential influence on player behavior in the Korean market. While developers optimize these strategies for spending and retention, their growing sophistication raises significant ethical concerns. As the mobile gaming industry continues to evolve, developers, regulators, and researchers must work collaboratively to address these challenges and promote ethical practices.

\renewcommand{\bibnumfmt}[1]{[#1]}%
\bibliographystyle{ACM-Reference-Format}
\bibliography{base}

\renewcommand{\bibnumfmt}[1]{[\citegameprefix#1]}%
\bibliographystylegame{ACM-Reference-Format}
\bibliographygame{games}

\end{document}